\documentstyle[preprint,tighten,aps,floats,epsfig]{revtex}
\def\ba{\begin{eqnarray}}
\def\ea{\end{eqnarray}}
\def\bq{\begin{equation}}
\def\eq{\end{equation}}
\def\lsim{\mathrel{\raisebox{-.6ex}{$\stackrel{\textstyle<}{\sim}$}}}
\def\gsim{\mathrel{\raisebox{-.6ex}{$\stackrel{\textstyle>}{\sim}$}}}

\begin{document}

\thispagestyle{empty}

\newcommand{\sla}[1]{/\!\!\!#1}

\preprint{
\font\fortssbx=cmssbx10 scaled \magstep2
\hbox to \hsize{
\hbox{\fortssbx University of Wisconsin - Madison}
\hfill\vtop{\hbox{\bf MADPH-00-1179}
            \hbox{May 2000}}}}

\title{\vspace*{.25in}
MEASUREMENT OF HIGGS PROPERTIES AT THE LHC\footnote{Talk given at the Rencontres de Moriond, {\it QCD and High Energy
Hadronic Interactions}, Les Arcs, France, March 18-25, 2000.}
}

\author{ D. ZEPPENFELD }

\address{Department of Physics, University of Wisconsin, Madison, WI 53706, 
USA\\
and\\
CERN, 1211 Geneva 23, Switzerland}

\maketitle
\begin{abstract}
A SM-like Higgs boson can be produced in a variety of channels at the LHC.
By combining information from production via gluon fusion and weak boson 
fusion, various partial widths and the total Higgs boson width can be 
extracted. Expected accuracies for 200~fb$^{-1}$ of data are in 
the 10\% range.
\end{abstract}

\section{Introduction}

One of the prime tasks of the LHC will be to probe the mechanism of 
electroweak gauge symmetry breaking. 
Beyond observation of the various CP even and CP odd scalars 
which nature may have in store for us,\cite{ATLAS,CMS}
this means the determination of 
the couplings of the Higgs boson to the known fermions and gauge bosons,
i.e. the measurement of $Htt$, $Hbb$, $H\tau\tau$ and $HWW$, $HZZ$, 
$H\gamma\gamma$ couplings, to the extent possible. 

Clearly this task very much depends on the expected Higgs boson mass. For
$m_H>200$~GeV and within the SM, only the $H\to ZZ$ and $H\to WW$ channels 
are expected to be observable, and the two gauge boson modes are related 
by SU(2). A much richer spectrum of decay modes
is predicted for the intermediate mass range, i.e. if a SM-like
Higgs boson has a mass between the reach of LEP2 ($\lsim 110$~GeV) and 
the $Z$-pair threshold. The main reasons for focusing on this range are 
present indications from electroweak precision
data, which favor $m_H<250$~GeV,\cite{EWfits} as well as expectations
within the MSSM, which predicts the lightest Higgs boson to have a mass
$m_h\lsim 130$~GeV. Recently, an analysis of Higgs coupling
measurements at the LHC was completed for this intermediate mass 
range~\cite{zknr} and in this talk I sumarize the results.

\section{Survey of intermediate mass Higgs channels}

The total production cross section for a SM Higgs boson at the LHC 
is dominated by the gluon fusion process, $gg\to H$, which largely 
proceeds via a top-quark loop. Thus, inclusive Higgs searches will 
collectively be called ``gluon fusion'' channels in the following. 
Three inclusive channels are highly promising for the SM Higgs boson 
search,\cite{ATLAS,CMS}
\ba 
\label{eq:ggHAA}
gg\to H\to \gamma\gamma\;, \qquad 
{\rm for}\quad m_H \lsim 150~{\rm GeV}\;, \\
\label{eq:ggHZZ}
gg\to H\to ZZ^*\to 4\ell\;,\qquad 
{\rm for}\quad m_H \gsim 130~{\rm GeV}\;, 
\ea
and
\bq
\label{eq:ggHWW}
gg\to H\to WW^*\to \ell\bar\nu\bar\ell\nu\;, \qquad 
{\rm for}\quad m_H \gsim 130~{\rm GeV}\;. 
\eq
The $H\to \gamma\gamma$ signal can be observed as a narrow and high 
statistics $\gamma\gamma$ invariant mass peak, albeit on a very large diphoton 
background. A few tens of $H\to ZZ^*\to 4\ell$ events are expected to be 
visible in 100~fb$^{-1}$ of data, with excellent signal to background 
ratios (S/B), ranging between 1:1 and 6:1, in a narrow four-lepton invariant 
mass peak. Finally, the $H\to WW^*\to \ell\bar\nu\bar\ell\nu$ mode is visible
as a broad enhancement of event rate in a 4-lepton transverse mass 
distribution, with S/B between 1:4 and 1:1 (for favorable values of the 
Higgs mass, around 170~GeV).

Additional, and, as we shall see, crucial information on the Higgs boson
can be obtained by isolating Higgs production in weak boson fusion (WBF), i.e.
by separately observing $qq\to qqH$ and crossing related processes, in which
the Higgs is radiated off a $t$-channel $W$ or $Z$.
Specifically, it was recently shown in parton 
level analyses that the weak boson fusion channels, with subsequent Higgs
decay into photon pairs,\cite{RZ_gamgam,R_thesis}
\bq
\label{eq:wbfHAA}
qq\to qqH,\;H\to \gamma\gamma\;, \qquad 
{\rm for}\quad m_H \lsim 150~{\rm GeV}\;,
\eq
into $\tau^+\tau^-$ pairs,\cite{R_thesis,RZ_tautau_lh,RZ_tautau_ll}
\bq
\label{eq:wbfHtautau}
qq\to qqH,\;H\to \tau\tau\;, \qquad 
{\rm for}\quad m_H \lsim 140~{\rm GeV}\;,
\eq
or into $W$ pairs~\cite{R_thesis,RZ_WW}
\bq
\label{eq:wbfHWW}
qq\to qqH,\;H\to WW^{(*)}\to e^\pm \mu^\mp /\!\!\!{p}_T\;, 
\qquad {\rm for}\quad m_H \gsim 120~{\rm GeV}\;,
\eq
can be isolated at the LHC. 
The weak boson fusion channels utilize the 
significant background reductions which are expected from 
double forward jet tagging~\cite{Cahn} and central jet vetoing 
techniques,\cite{bjgap} and promise low
background environments in which Higgs decays can be studied in detail.

\begin{table*}[t]
\vspace{0.2in}
\caption{Number of events expected for 
$qq\to qqH,\;H\to WW^{(*)}\to\mu^\pm e^\mp\sla p_T$ 
in 200~fb$^{-1}$ of data, and corresponding 
backgrounds.\protect\cite{RZ_WW}
The expected relative statistical error on the signal cross section 
is given in the last line.}
\vspace{0.15in}
\label{table:WBF.WW}
\begin{tabular}{c|cccccccc}
$m_H$ &  120  &  130  &  140  &  150  &  160  &  170  &  180  &  190 \\
\hline
$N_S$ &  136  &  332  &  592  &  908  & 1460  & 1436  & 1172  &  832 \\
$N_B$ &  136  &  160  &  188  &  216  &  240  &  288  &  300  &  324 \\
$\Delta\sigma_H/\sigma_H$ &
       12.1\% & 6.7\% & 4.7\% & 3.7\% & 2.8\% & 2.9\% & 3.3\% & 4.1\%  \\
\end{tabular}
\end{table*}

\begin{figure}[thb]
\centering\leavevmode
\epsfig{file=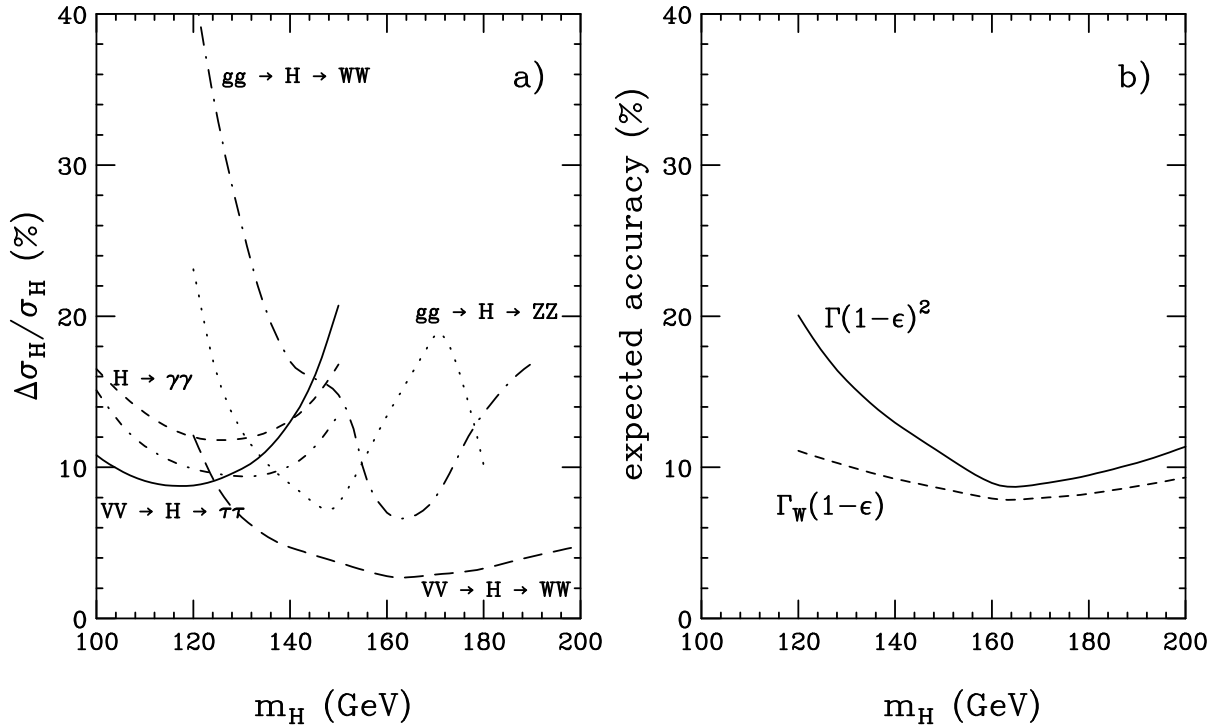,width=9.7cm,angle=90}
\vspace*{0.2cm}

\caption{Relative accuracy expected at the LHC with 200~fb$^{-1}$ of data.
a) Cross section times branching fraction for several inclusive modes 
(dotted and dash-dotted lines) and WBF channels (dashed and solid lines).
b) Extracted total width (solid line) and $H\to WW$ partial width 
(dashed line). }
\label{fig1}
\end{figure}

An example of expected events rates (after cuts and including efficiency 
factors) are summarized in Table~\ref{table:WBF.WW} for the 
$qq\to qqH,\;H\to WW^{(*)}\to e^\pm \mu^\mp\sla p_T$ signal. 
The rates and ensuing statistical errors of the signal cross section 
are given for 100~fb$^{-1}$ of data collcted in both the ATLAS and the 
CMS detector. 
The statistical accuracy with which the signal cross sections of the 
processes in Eqs.~(\ref{eq:ggHAA}-\ref{eq:wbfHWW}) can be determined is
shown in Fig.~\ref{fig1}a).

\section{Measurement of Higgs properties}
\label{sec3}

In order to translate the cross section measurements of the various
Higgs production and decay channels into measurements of Higgs boson 
properties, in particular into measurements of the various Higgs boson
couplings to gauge fields and fermions, it is convenient to rewrite 
them in terms of partial widths of various Higgs boson decay channels. 
The Higgs-fermion couplings $g_{Hff}$, for example, which in the SM are 
given by the fermion masses, $g_{Hff} = m_f(m_H)/v$, can be traded for 
$\Gamma_f = \Gamma(H\to \bar ff)$. Similarly the 
square of the $HWW$ coupling ($g_{HWW}=gm_W$ in the SM) or the $HZZ$ 
coupling is proportional to the partial widths $\Gamma_W=\Gamma(H\to WW^*)$ 
or $\Gamma_Z=\Gamma(H\to ZZ^*)$.
$\Gamma_\gamma=\Gamma(H\to\gamma\gamma)$ and $\Gamma_g=\Gamma(H\to gg)$
determine the squares of the effective $H\gamma\gamma$ and $Hgg$ couplings.
The Higgs production cross sections are governed by the same squares of
couplings, hence, $\sigma(VV\to H) \sim \Gamma_V$ (for $V=g,\;W,\;Z$).
Combined with the branching fractions $B(H\to ii)=\Gamma_i/\Gamma$
the various signal cross sections measure different combinations of Higgs
boson partial and total widths, $\Gamma_i\Gamma_j/\Gamma$. 

The production rate for WBF is a mixture of $ZZ\to H$ and $WW\to H$ 
processes, and we cannot distinguish between the two 
experimentally. In a large class of models the ratio of $HWW$ and $HZZ$ 
couplings is identical to the one in the SM, however, and this includes the
MSSM. We therefore assume that 1) the $H\to ZZ^*$ and $H\to WW^*$ partial 
widths are related by SU(2) as in the SM, i.e. their ratio, $z$, is given 
by the SM value, $z=\Gamma_Z/\Gamma_W=z_{SM}$.
Note that this assumption can be tested, at the 15-20\% level for 
$m_H>130$~GeV, by forming the ratio 
$B\sigma(gg\to H\to ZZ^*)/B\sigma(gg\to H\to WW^*)$.

With  $W,Z$-universality, the three weak boson fusion cross sections give us
direct measurements of three combinations of (partial) widths,
\ba
X_\gamma = {\Gamma_W\Gamma_\gamma\over \Gamma}\qquad &{\rm from}&\;\;
qq\to qqH,\; H\to\gamma\gamma \;, \\
X_\tau = {\Gamma_W\Gamma_\tau\over \Gamma}\qquad &{\rm from}&\;\;
qq\to qqH,\; H\to\tau\tau \;, \\
X_W = {\Gamma_W^2\over \Gamma}\quad\qquad &{\rm from}&\;\;
qq\to qqH,\; H\to WW^{(*)} \;,
\ea
In addition the three gluon fusion channels provide measurements of 
\ba
Y_\gamma = {\Gamma_g\Gamma_\gamma\over \Gamma}\qquad &{\rm from}&\;\;
gg\to H\to\gamma\gamma \;, \\
Y_Z = {\Gamma_g\Gamma_Z\over \Gamma}\qquad &{\rm from}&\;\;
gg\to H\to ZZ^{(*)} \;, \\
Y_W = {\Gamma_g\Gamma_W\over \Gamma}\qquad &{\rm from}&\;\;
gg\to H\to WW^{(*)} \;.
\ea
When extracting Higgs couplings, the QCD uncertainties of production cross
sections enter. These can be estimated via the residual scale dependence
of the NLO predictions and are small (of order 5\%) for the WBF 
case,\cite{wbfNLO} while larger uncertainties of about 20\% are found 
for gluon fusion.\cite{HggNLO}

A first test of the Higgs sector is provided by taking ratios 
of the $X_i$'s and ratios of the $Y_i$'s. QCD uncertainties, and all 
other uncertainties related to the initial state, like luminosity and pdf 
errors, cancel in these ratios.  They test $W,Z$-universality,
and compare the $\tau\tau H$ Yukawa coupling with the $HWW$ coupling.
Typical errors on these cross 
section ratios are expected to be in the 15 to 20\% range.
Accepting an additional systematic error of about 20\%, a measurement 
of the ratio $\Gamma_g/\Gamma_W$, which determines the $Htt$ to $HWW$ 
coupling ratio, can be performed, by measuring the cross section 
ratios $B\sigma(gg\to H\to\gamma\gamma)/\sigma(qq\to 
qqH)B(H\to\gamma\gamma)$
and  $B\sigma(gg\to H\to WW^*)/\sigma(qq\to qqH)B(H\to WW^*)$. 

Beyond the measurement of coupling ratios, minimal additional assumptions 
allow an indirect measurement of the total Higgs width. First of all, the 
$\tau$ partial width, properly normalized, is measurable with an accuracy of 
order 10\%. The $\tau$ is a third generation fermion with isospin 
$-{1\over 2}$, just like the $b$-quark. In many models, the ratio of their
coupling to the Higgs is given by the $\tau$ to $b$ mass ratio. 
In addition to $W,Z$-universality we thus assume that
(2) $y = \Gamma_b/ \Gamma_\tau = y_{SM}$ and, finally,
(3) the branching ratio for unexpected channels is small, i.e.
$\epsilon = 1-(B(H\to b\bar b)+B(H\to \tau\tau)+B(H\to WW^{(*)})+
B(H\to ZZ^{(*)})+B(H\to gg)+B(H\to \gamma\gamma) \ll 1$.

With these three assumptions consider the observable
\begin{eqnarray}
\tilde\Gamma_W &=& X_\tau(1+y) + X_W(1+z) + X_\gamma + Y_W 
\nonumber \\
&=& \biggl(\Gamma_\tau+\Gamma_b +\Gamma_W +\Gamma_Z+
\Gamma_\gamma+\Gamma_g\biggr){\Gamma_W\over\Gamma}
=(1-\epsilon)\Gamma_W  \;.
\end{eqnarray}
$\tilde\Gamma_W$ provides a lower 
bound on $\Gamma(H\to  WW^{(*)})=\Gamma_W$. Provided $\epsilon$ is small 
(within the SM and for $m_H>110$~GeV, $\epsilon < 0.04$ and it is dominated 
by $B(H\to c\bar c)$), the determination 
of $\tilde\Gamma_W$ provides a direct measurement of the $H\to WW^{(*)}$ 
partial width. Once $\Gamma_W$ has been determined, the total width of the 
Higgs boson is given by
\bq
\label{eq:Gamma_tot}
\Gamma = {\Gamma_W^2\over X_W}={1\over X_W}
\biggl(X_\tau(1+y) + X_W(1+z) + X_\gamma + \tilde X_g \biggr)^2
{1\over (1-\epsilon)^2}\; .
\eq

The extraction of the total Higgs width, via Eq.~(\ref{eq:Gamma_tot}),
requires a measurement of the $qq\to qqH, H\to WW^{(*)}$ cross section,
which is expected to be available for $m_H\gsim 115$~GeV.\cite{RZ_WW}
Consequently, errors are large for Higgs masses close to this lower limit
but decrease to about 10\% for Higgs boson masses around the $WW$ threshold.
Results are shown in Fig.~\ref{fig1}b) and look highly promising.

\section{Summary}
\label{sec4}

With an integrated luminosity of 100 fb$^{-1}$ per experiment, the LHC can 
measure various ratios of Higgs partial widths, with accuracies of order 
10 to 20\%. This translates into 5 to 10\%
measurements of various ratios of coupling constants. The 
ratio $\Gamma_\tau/\Gamma_W$ measures the coupling of down-type fermions
relative to the Higgs couplings to gauge bosons. To the extent that the
$H\gamma\gamma$ triangle diagrams are dominated by the $W$ loop,
the width ratio $\Gamma_\tau/\Gamma_\gamma$ probes the same relationship.
The fermion triangles leading to an effective $Hgg$ coupling are expected 
to be dominated by the top-quark, thus, $\Gamma_g/\Gamma_W$ probes the 
coupling of up-type fermions relative to the $HWW$ coupling.
Finally, for Higgs boson masses above $\approx 120$~GeV,
the absolute normalization of the $HWW$ coupling is accessible
via the extraction of the $H\to WW^{(*)}$ partial width
in weak boson fusion.

These measurements test the crucial aspects of the Higgs sector.
The $HWW$ coupling, being linear in the Higgs field, 
identifies the observed Higgs boson as the scalar 
responsible for the spontaneous breaking of $SU(2)\times U(1)$: a 
scalar without a vacuum expectation value does not exhibit such a 
trilinear coupling at tree level. 
The measurement of the ratios of $g_{Htt}/g_{HWW}$ and 
$g_{H\tau\tau}/g_{HWW}$ then probes the mass generation of both up and down
type fermions. Thus the LHC can do much more than mereley discover the Higgs:
it can give us detailed and reasonably precise information on the dynamics 
of electroweak symmetry breaking.

\section*{Acknowledgments}
I would like to thank R.~Kinnunen, A.~Nikitenko and E.~Richter-Was for a 
most enjoyable collaboration leading to the results summarized here, and the 
CERN theory group for its hospitality.
This work was supported in part by the University of Wisconsin Research
Committee with funds granted by the Wisconsin Alumni Research Foundation and
in part by the U.~S.~Department of Energy under Contract
No.~DE-FG02-95ER40896. 


\end{document}